\documentclass[useAMS,usegraphicx,usenatbib]{mn2e} 
\usepackage{array}
\usepackage{rotating}
\bibpunct{(}{)}{;}{a}{}{,}
\def\logg{\log g}
\def\vlos{v_{\rm los}}
\def\gta{{_ >\atop{^\sim}}}

\newcolumntype{M}[1]{>{\raggedright}m{#1}}

\title[A radial velocity gradient in the Galactic disc]{Detection of a radial velocity gradient in the extended local disc with RAVE}

\author[A.    Siebert   et    al.]    {A.~Siebert$^{1}$\thanks{E-mail:
arnaud.siebert@astro.unistra.fr},  B.~Famaey$^{1}$,  I.~Minchev$^{1}$,
G.M.~Seabroke$^{2}$,   J.~Binney$^{3}$,  B.~Burnett$^{3}$,  \newauthor
K.C.~Freeman$^{4}$,       M.~Williams$^{5}$,      O.~Bienaym\'e$^{1}$,
J.~Bland-Hawthorn$^{6}$,         R.~Campbell$^{5}$,         \newauthor
J.P.~Fulbright$^{7}$,       B.K.~Gibson$^{8}$,       G.~Gilmore$^{9}$,
E.K.~Grebel$^{10}$,   A.~Helmi$^{11}$,   U.~Munari$^{12}$,  \newauthor
J.F.~Navarro$^{13}$,       Q.A.~Parker$^{14}$,       W.A.~Reid$^{14}$,
A.~Siviero$^{5,12}$,  M.~Steinmetz$^{5}$,  F.~Watson$^{15}$,  \newauthor
R.F.G.~Wyse$^{7}$,       T.~Zwitter$^{16,17}$\\       $^1$Observatoire
Astronomique, Universit\'e de  Strasbourg, CNRS, Strasbourg, France \\
$^2$Mullard  Space  Science  Laboratory,  University  College  London,
Holmbury St  Mary, Dorking, RH5  6NT, UK \\ $^3$Rudolf  Peierls Centre
for   Theoretical  Physics,   Oxford,  UK\\   $^4$Australian  National
University,  Canberra,   Australia\\  $^5$Astrophysikalishes  Institut
Potsdam  ,  Potsdam,  Germany\\  $^6$Sydney Institute  for  Astronomy,
University   of  Sydney,  Sydney,   Australia  \\   $^7$Johns  Hopkins
University, Baltimore, MD, USA\\ $^8$University of Central Lancashire,
Preston,   UK\\   $^9$Institute    of   Astronomy,   Cambridge,   UK\\
$^{10}$Astronomisches  Rechen-Institut, Zentrum  f\"ur  Astronomie der
Universit\"at   Heidelberg,   Heidelberg,   Germany\\   $^{11}$Kapteyn
Astronomical      Institut,      Groningen,     the      Netherlands\\
$^{12}$Astronomical Observatory  of Padova in  Asiago, Asiago, Italy\\
$^{13}$University  of  Victoria,  Victoria,  Canada\\  $^{14}$Macquary
University,  Sydney, Australia\\  $^{15}$Anglo-Australian Observatory,
Sydney,  Australia\\   $^{16}$University  of  Ljubljana,   Faculty  of
Mathematics  and  Physics,   Ljubljana,  Slovenia\\  $^{17}$Center  of
excellence, SPACE-SI, Ljubljana, Slovenia }

\begin{document}

\date{Accepted . Received ; in original form \today}

\maketitle

\begin{abstract}  
Using a sample of 213\,713  stars from the Radial Velocity Experiment (RAVE)
survey, limited to  a distance of 2~kpc from the Sun  and to $|z|<1$~kpc, we
report the  detection of  a velocity  gradient of disc  stars in  the fourth
quadrant, directed radially  from the Galactic centre.  In  the direction of
the Galactic centre, we apply  a simple method independent of stellar proper
motions and of Galactic parameters  to assess the existence of this gradient
in the  RAVE data.  This velocity  gradient corresponds to $|K+C|  \gta 3 \;
{\rm km} \, {\rm s}^{-1} \, {\rm  kpc}^{-1}$, where $K$ and $C$ are the Oort
constants measuring  the local divergence  and radial shear of  the velocity
field, respectively.   In order  to illustrate the  effect, assuming  a zero
radial  velocity of  the  Local Standard  of  Rest we  then reconstruct  the
two-dimensional  Galactocentric velocity  maps using  two different  sets of
proper motions  and photometric distances based either  on isochrone fitting
or  on $K$-band  magnitudes,  and considering  two  sets of  values for  the
Galactocentric  radius  of  the  Sun  and  local  circular  speed.   Further
observational confirmation  of our finding with  line-of-sight velocities of
stars  at  low  latitudes,  together  with further  modelling,  should  help
constrain  the  non-axisymmetric   components  of  the  Galactic  potential,
including the bar, the spiral arms  and possibly the ellipticity of the dark
halo.
\end{abstract}


%
%
\section{Introduction}

Our  Milky Way  Galaxy is  a unique  laboratory in  which to  study galactic
structure and evolution from stellar kinematics. The story of the efforts to
obtain such  kinematic data nicely illustrates how  theoretical progress and
data acquisition have to  go hand in hand if one wants  to gain insight into
the structure and history of  the Galaxy.  Until recently, most studies have
been limited to the solar neighbourhood.  The zeroth order approximation for
modelling these kinematic  data assumes an axisymmetric model,  in which the
Local Standard of Rest (LSR) is  on a perfectly circular orbit. However, the
local velocity  field in the  solar neighbourhood already  displays possible
signatures of the  non-axisymmetry of the Galactic potential  in the form of
stellar moving groups  containing stars of very different  ages and chemical
compositions \citep{dehnen1998,fetal05,quillen2005,bensby07,minchev2010}.

In  external  galaxies,   large-scale  distortions  of  the  (line-of-sight)
isovelocity contours of the gas have long been recognised in two-dimensional
velocity  maps \citep[e.g.][]{bosma1978}, allowing  a quantification  of the
strength of non-circular streaming motions \citep[e.g.,][]{sellwood2010}.  A
prominent example  is the  galaxy M81 \citep[e.g.,][]{adler1996},  which has
roughly the same circular velocity as  the Milky Way and in which the radial
streaming motions  of the gas reach  maxima of the order  of 50~km s$^{-1}$.
However, the stellar  kinematics often turn out to  be slightly more regular
and  symmetric  than the  gas  kinematics \citep[e.g.][]{pizzella2008}.  For
instance, \citet{rix95}  estimate from 18  face-on spiral galaxies  that the
azimuthally averaged radial streaming motions of disk stars should typically
amount  to a  maximum  of 7~$\mathrm{km}\,\mathrm{s}^{-1}$  due to  lopsided
distortions   and    6~$\mathrm{km}\,\mathrm{s}^{-1}$   due   to   intrinsic
ellipticity.\\

In  the   Milky  Way,  non-axisymmetric  motions  are   clearly  visible  in
radio-frequency observations of gas flow  in the inner Galaxy, implying that
the amplitude of spiral structure could be larger by a factor $\sim$1.5 than
its     amplitude     in     the    near-infrared     luminosity     density
\citep{bissantz2003,famaey2005}.   On the  other hand,  radio interferometry
used to determine the parallax of star-forming regions in the Perseus spiral
arm, together with the  measurement of their line-of-sight (los) velocities,
show  that these  star-forming  regions have  peculiar non-circular  motions
amounting to $\sim$20~km~s$^{-1}$, assuming standard IAU Galactic parameters
and   a  nearly-flat   rotation   curve  \citep{xu2006,binney2006,reid2009}.
However, the amplitude of such  non-circular motions strongly depends on the
Galactic parameters adopted  \citep{mcmillan2010}, namely the Galactocentric
radius of the  Sun $R_0$, the local circular speed  $v_{c0}$, and the reflex
solar motion, as well as the  model for the rotation curve.  Another example
of  such  non-circular  motion  is  the 3~kpc  arm  which  has  $\vlos=\,-80
\mathrm{km}\,\mathrm{s}^{-1}$ towards the Galactic  centre (GC) and is known
to  have a  counterpart on  the far  side of  the GC  \citep{dame2009}.  Its
origin is associated to the Galactic bar showing that non-circular motion of
the similar  amplitude can be caused  by various types  of perturbations and
not necessarily by spiral arms.\\

Here, we  use line-of-sight velocity\footnote{Throughout this  paper we will
  use  the  notation $\vlos$  or  line-of-sight  velocity  to refer  to  the
  heliocentric radial velocity measured  by RAVE while {\it radial} velocity
  or $V_R$  means {\it galactocentric radial} velocity}  data ($\vlos$) from
the RAVE survey to investigate whether  the mean radial motion of disc stars
remains zero in  the extended solar neighbourhood as compared  to the one in
the solar neighbourhood, or whether large-scale radial streaming motions are
present  in the  vicinity of  the  Sun, which  would translate  in a  radial
velocity gradient in  one or the other direction.  In  Sec.~2 we present the
sample used for  our analysis while in Sec.~3 we present  the detection of a
radial velocity gradient using RAVE  $\vlos$, as well as the associated Oort
constants and  2-dimensional velocity field.   Finally in Sec.~4  we discuss
the implications of our results and present our conclusions.

%
%
\section{The RAVE sample}
\label{s:sample}

Our analysis is  based on the RAVE catalogue  \citep{dr1,dr2} which provides
$\vlos$   to   2~$\mathrm{km}\,\mathrm{s}^{-1}$   and  stellar   atmospheric
parameters and distances  to 30\% for a large number of  bright stars in the
southern  hemisphere   with  $9<I<12$  using   moderate  resolution  spectra
(R$\sim$7\,500).   RAVE  selects  its   targets  randomly  in  the  $I$-band
interval, and so  its properties are similar to  a magnitude limited survey.
The RAVE catalogue is  cross-matched with astrometric (PPMX, UCAC2, Tycho-2)
and  photometric  catalogues (2MASS,  DENIS)  to  provide additional  proper
motions and magnitudes making RAVE a suitable tool to study the Milky Way.

We  use  the latest  version  of  the  \citet{zwitter2010} catalogue,  which
contains  four sets  of  distances in  addition  to the  standard RAVE  data
(line-of-sight  velocities  and  stellar  parameters). These  four  sets  of
distances  are  (i)  the  distance  moduli  computed  using  three  sets  of
isochrones  (including the  Padova isochrones),  and (ii)  the  distances of
\citet{breddels2010}.  In this  paper we will use the  distance moduli based
on the Padova  isochrones but our results do not strongly  depend on the set
of  isochrones used  (see  also Sect.\ref{s:oort}  where  all distances  are
multiplied  by  an  arbitrary  factor).   This  catalogue  is  cleaned  from
potentially problematic  fields using a list kindly  provided by G.~Matijevic
(private  communication) and  we further  select stars  with  $|z|<1$~kpc to
minimise  contamination by  halo stars.   This leaves  us with  a  sample of
213\,713  stars  with known  distances,  atmospheric  parameters ($\log  g$,
$T_\mathrm{eff}$, $[m/H]$),  proper motions as well  as $\vlos$ measurements
that can be used to compute the 3D space velocities.

\begin{figure}
\includegraphics[width=7cm]{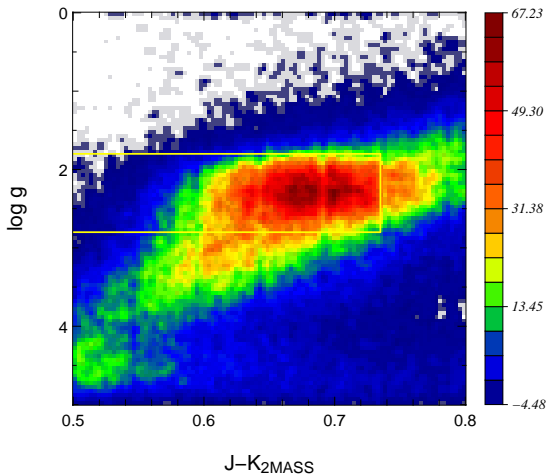}
\caption{Distribution of  stars in the  ($\logg$, $J-K$) plane near  the red
  clump for the  RAVE internal release. Colour coding  follows the number of
  stars per  bin of 0.03~mag  width in $J-K$  and 0.05~dex in  $\logg$.  The
  distribution is smoothed  over 3 pixels to reduce the  noise. Stars in the
  white box are identified as red-clump stars.}
\label{f:sample}
\end{figure}

\begin{figure*}
\begin{tabular}{c c}
{\bf Full sample} & {\bf Red clump candidates}\\
\includegraphics[width=6cm]{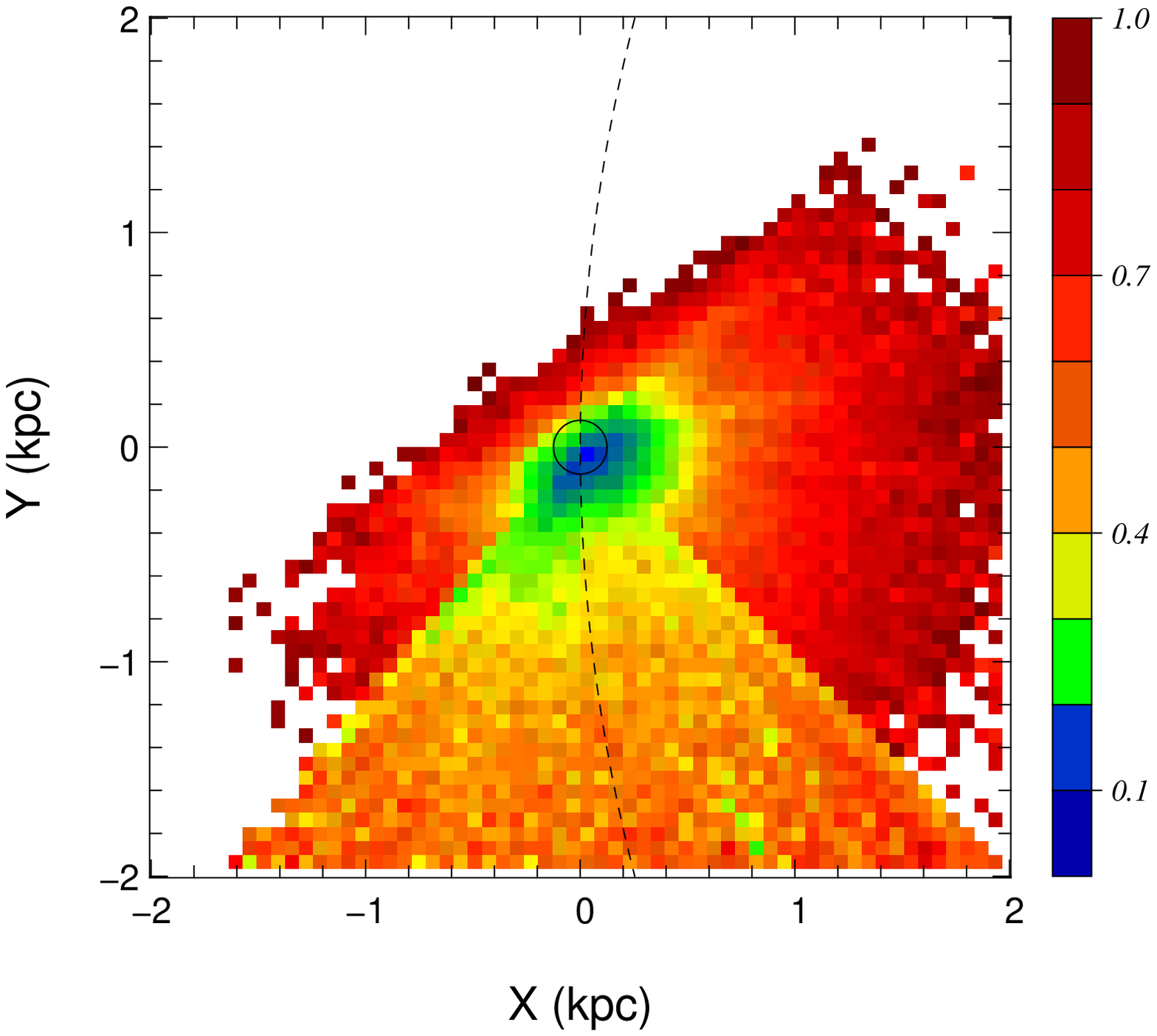}&
\includegraphics[width=6cm]{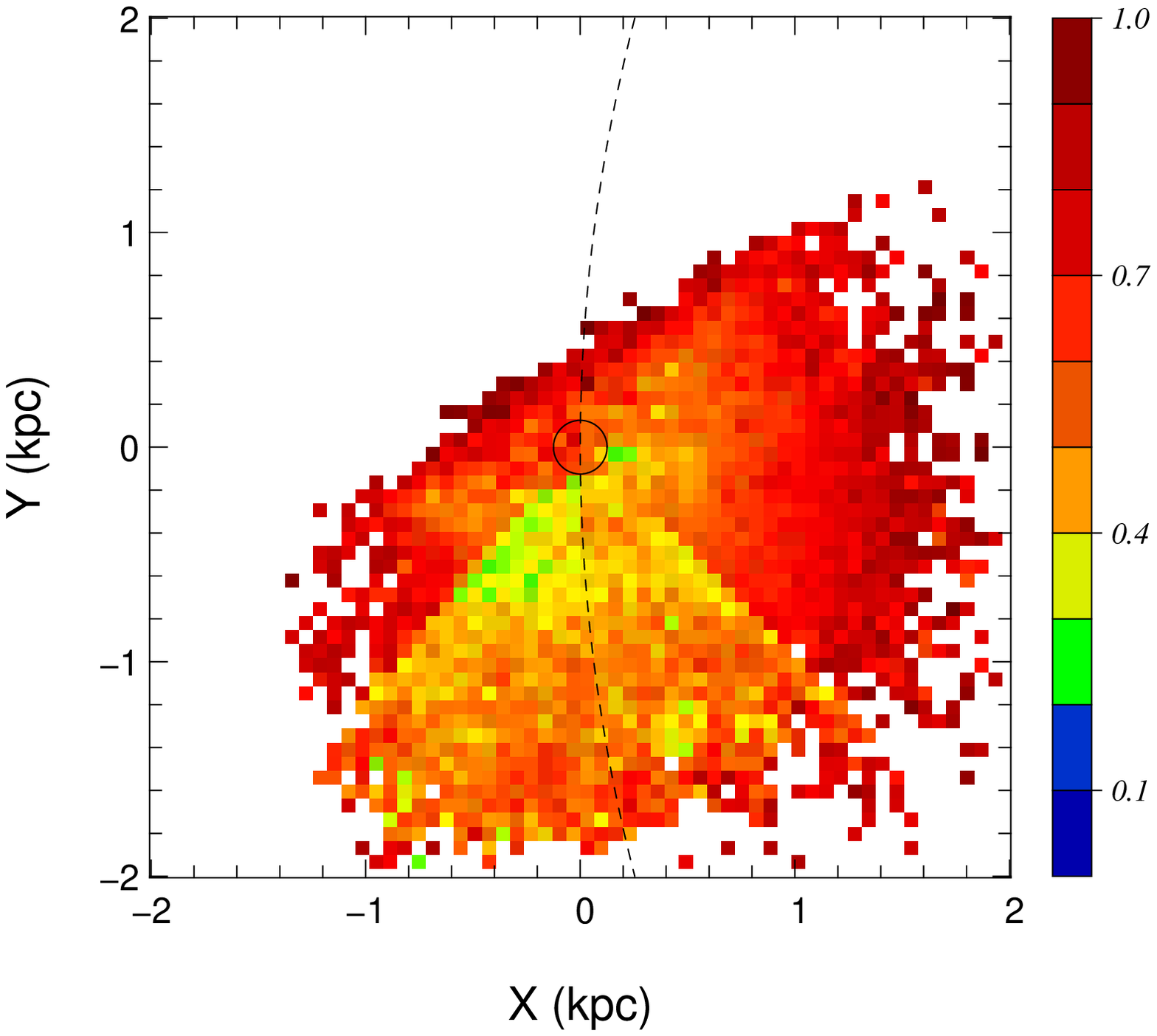}\\
\end{tabular}
\caption{Mean  $|z|$ distance  to  the Galactic  plane  in kpc  for the  two
  samples studied:  full sample (left),  red clump sample (right).   The two
  panels  contain $60\times60$  pixels in  a box  $4\times4$~kpc  in extent.
  Both samples have  an upper limit on $|z|$,  $max(|z|)=1$~kpc.  The Sun is
  located at the centre of the map,  X is increasing in the direction of the
  Galactic centre and Y towards  the Galactic rotation.  The solar circle is
  shown as  a dashed  line.  The  circle in the  centre represents  a sphere
  125~pc in radius.}
\label{f:z}
\end{figure*}

To verify that our analysis does  not depend on the method used to determine
the distances  or on  the source of  proper motions,  we also use  stars for
which distances and proper motions are obtained independently.  To do so, we
select a  sub-sample of red clump  candidates for which the  distance can be
estimated  using the  $K$--magnitude  alone because  of  the narrow,  almost
Gaussian, luminosity  function of the  red clump.  Red clump  candidates are
selected   in    the   ($\logg$,    $J-K$)   plane,   as    illustrated   in
Fig.~\ref{f:sample}.  We  select the over-density  using $0.5<J-K<0.735$ and
$1.8<\logg<2.8$: the red  limit helps reduce the contamination  by the upper
red-giant-branch stars.   Distances are obtained  assuming all stars  in the
sample are  red clump  stars using a  Gaussian luminosity function  for this
population with $M_K=-1.6$ and $\sigma_K=0.22$. We apply the same cut in $z$
as for  the full  sample. This  results in a  sample containing  29\,623 red
clump candidates. More details on  the contamination and distance errors can
be found  in \citet{siebert08}.  We  cross-match this sample with  the UCAC3
catalogue to  provide a different source  for the proper motion  -- the RAVE
catalogue contains  predominantly Tycho-2  and PPMX proper  motions.  Hence,
for this sample,  the distances and proper motions  are independent from the
full RAVE sample.   The mean height above/below the  Galactic plane for each
sample is shown in Fig.~\ref{f:z}.

%
%
\section{A radial velocity gradient}

\subsection{Line-of-sight velocity detection}
\label{s:RV}

Most of the RAVE stars lie in the fourth Galactic quadrant with the tails of
the  distribution in  Galactic longitude  extending in  the first  and third
quadrants. For our analysis to remain independent of Galactic parameters and
proper  motions, we first  look in  the GC  (and anticentre)  direction with
$|l|<5\degr$ (and $175\degr<l<185\degr$).   In this case, the Galactocentric
radial velocity is close to the measured line-of-sight velocity projected on
the plane (or  its opposite in the anticentre direction),  with only a small
contribution from the proper motion vector.

\begin{figure*}
\begin{tabular}{c c}
{\bf \hspace{1cm} Full sample} & {\bf  \hspace{1cm} Red clump candidates}\\
\includegraphics[width=6cm]{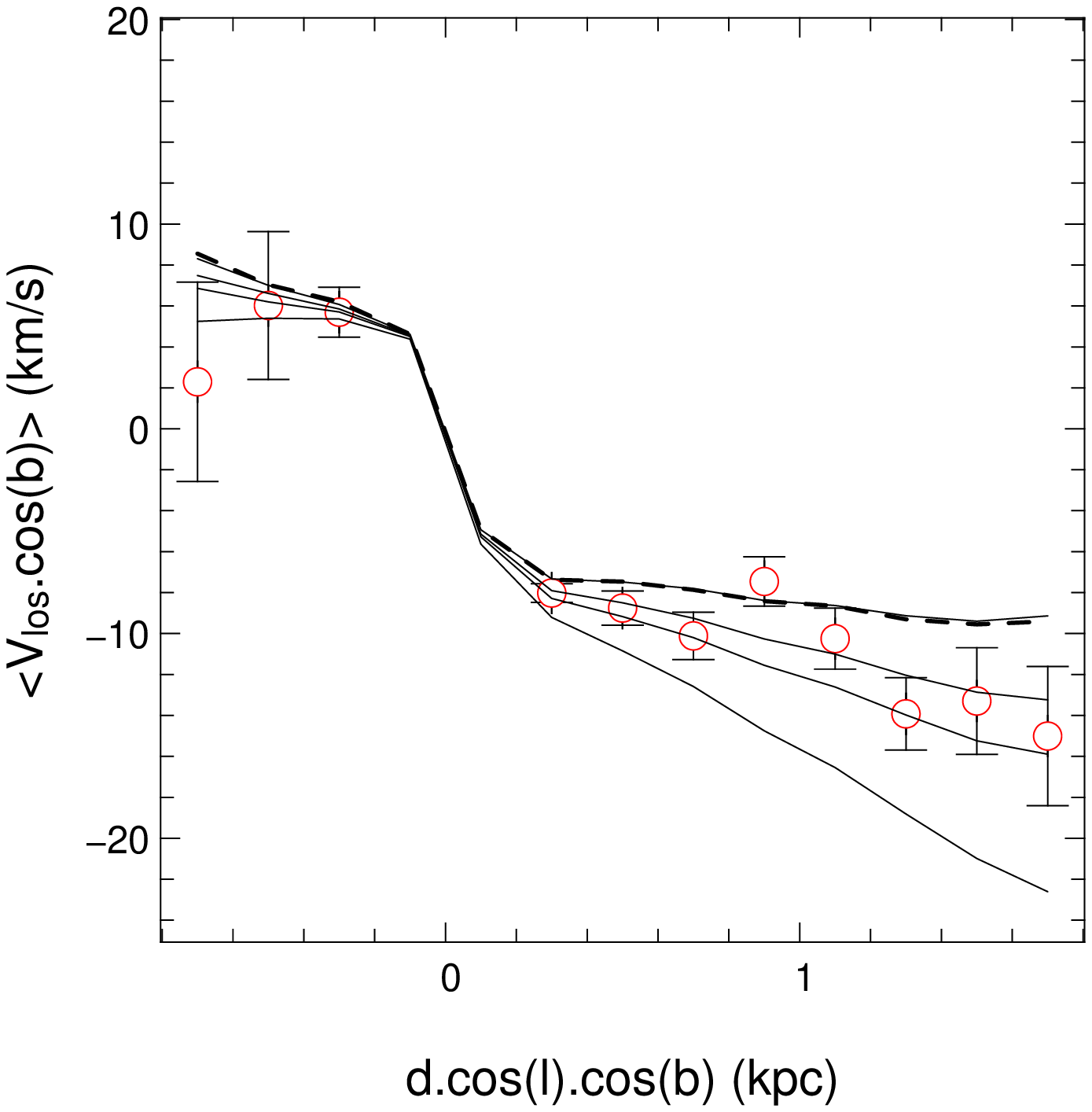}&
\includegraphics[width=6cm]{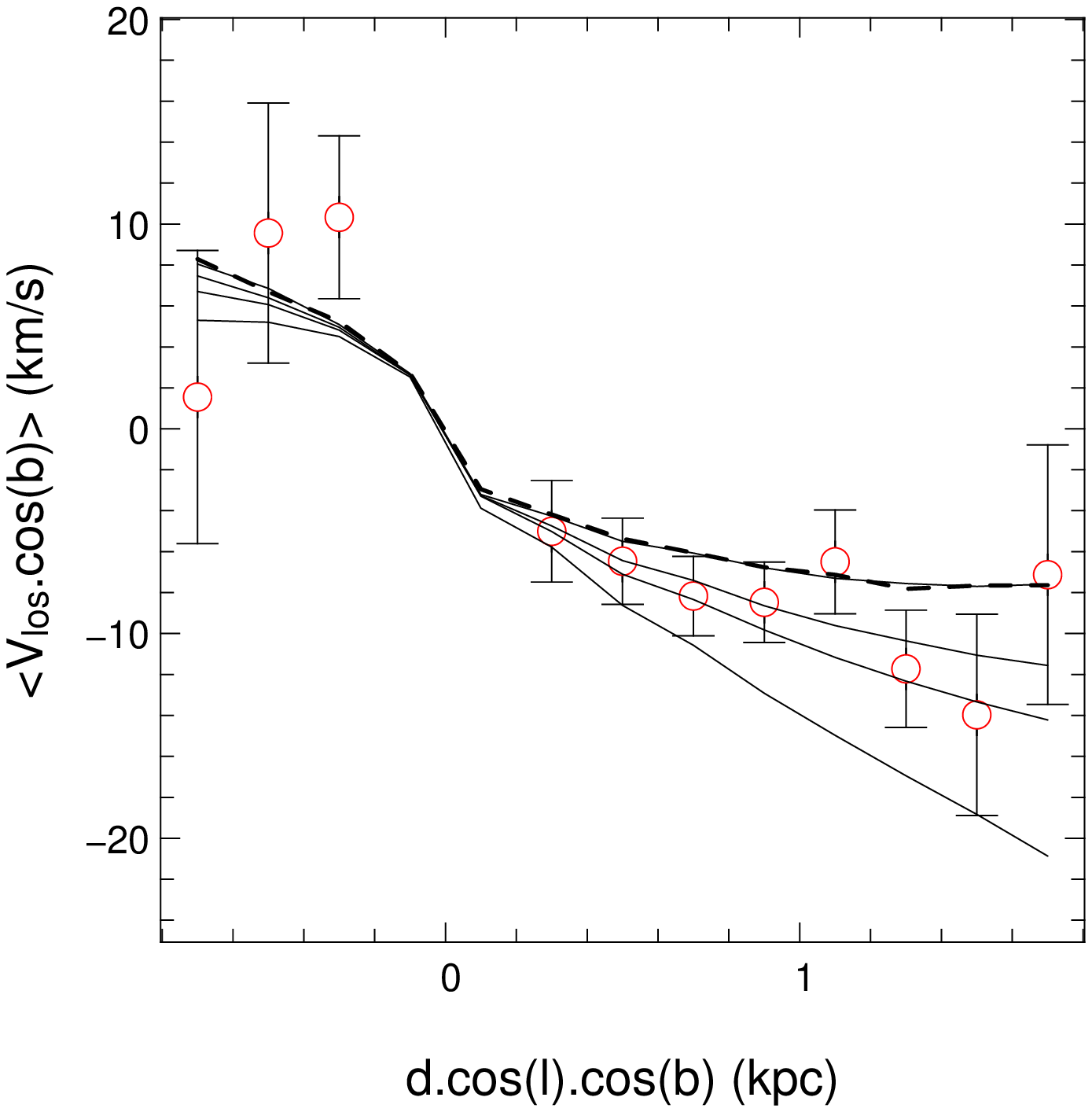}\\
\end{tabular}
\caption{Projection  of the  mean  RAVE  $\vlos$ on  the  Galactic plane  in
  distance  intervals of  200~pc in  the  direction of  the Galactic  centre
  ($|l|<5  \degr$)  and   anticentre  ($175\degr<l<185\degr$)  for  our  two
  samples.  Left:  full sample containing  10\,682 stars.  Right:  red clump
  sample  containing 1\,328  stars.   The  open circles  are  our data  with
  $1-\sigma$ error bars. The dashed  curve corresponds to a thick disc model
  with no net radial motion, while the  full curves are for a thin disc with
  a radial velocity  gradient of 0, 3, 5 and 10  km s$^{-1}$ kpc$^{-1}$ from
  top to bottom.}
\label{f:rvonly_l0}
\end{figure*}

In Fig.~\ref{f:rvonly_l0} we plot the  projection onto the plane of the mean
$v_{\rm los}$  in bins $200$~pc wide as  a function of $d\cos  \ell \cos b$.
The left panel is for the full  sample and the right panel for the red clump
candidates.  We  then compare the  observed mean velocities to  the expected
velocities for  a thin disc  in circular rotation  at $v_{c0}$ and  adding a
radial gradient  $\partial \langle V_R  \rangle/\partial R=$3, 5,  and 10~km
s$^{-1}$ kpc$^{-1}$, where $V_R$  is the Galactocentric radial velocity (see
full curves  from top  to bottom in  each panel  in Fig.~\ref{f:rvonly_l0}).
The curves are  drawn using 5\,000 Monte Carlo  realisations of the samples,
replacing  the measured  velocities by  random realisations  of a  thin disc
ellipsoid. The model used is the same for both samples, kinematic parameters
for  the  thin disc  ellipsoid  being  taken  from \citet{BM98}  table~10.4.
Clearly,  our data  are  not compatible  with  a disc  in circular  rotation
because the  mean projected  line-of-sight velocity is  systematically lower
than the expected mean velocity.  The offset between the two panels reflects
the different distribution in $R$ of the two samples.

Because our samples are not  perfectly oriented towards the GC, the rotation
component  could bias  our  measurement.  Therefore,  we  reproduce this  by
modelling a thick  disc population which lags the  circular rotation by 36~km
s$^{-1}$. The result  for a thick disc is presented as  dashed curves and is
computed only in  the case of circular rotation.  The  results do not differ
significantly from  the thin disc  results indicating that  the contribution
from $v_{c0}$ to our line-of-sight velocities is marginal and hence that our
result truly measures a radial velocity gradient in the RAVE data.

%
%

\subsection{Oort constants}
\label{s:oort}

As first  worked out by  Oort (1927) for  an axisymmetric Galaxy,  and later
generalised to  non-axisymmetry (e.g., Chandrasekhar 1942),  a practical way
to study  the global  streaming motion of  field stars  in the Galaxy  is to
consider an arbitrary smooth velocity  field and Taylor expand it around the
position  of the Sun.   Such a  Taylor expansion  of the  Cartesian velocity
components ($U$,$V$) to first order yields  the four Oort constants A, B, C,
and  K,  measuring  the  azimuthal  (A)  and radial  (C)  shear,  the  local
divergence  (K), and  the  vorticity (B)  of  the velocity  field (see  e.g.
Kuijken \&  Tremaine 1994,  Olling \& Dehnen  2003, hereafter KT94  and OD03
respectively).

While we note that such a first  order Taylor expansion might not be a valid
description of  the velocity field for  sample depths larger  than 1~kpc (as
already noted in Pont et al. 1994  and OD03), we also note that, at least in
the direction of  the GC, a constant linear  velocity gradient $\partial V_R
/\partial R=K+C$ between $-3$ and $-10$ km s$^{-1}$ kpc$^{-1}$ is sufficient
to reproduce  the trend seen in  Fig.\ref{f:rvonly_l0}. In view  of this, we
compute the Oort  constants in the classical way using

\begin{eqnarray}
\langle v_{rad}/r \rangle=K+A\sin2\ell+C\cos2\ell\,\, , \nonumber\\
\langle v_t/r \rangle=B+A\cos 2\ell- C\sin 2\ell\,\, ,
\label{e:oort}
\end{eqnarray}

where   $v_{rad}$  (resp.    $v_t$)  is   the  heliocentric   radial  (resp.
transverse) component of the velocity vector in the Galactic plane corrected
for the motion of the Sun with  respect to the LSR. $r$ is the distance from
the star to  the Sun in the Galactic  plane and $A$ $B$ $C$ and  $K$ are the
Oort constants  we try to  measure (see also  KT94, OD03).  We then  fit the
right-hand side of eq.~\ref{e:oort} to the left-hand side that is built upon
our observations with $v_{rad}$ depending on $\vlos$ and $\mu_b$ while $v_t$
depends on $\mu_\ell$.

We restrict the sample to  the $-140<\ell<10\deg$ interval, the noise in the
computed  mean  velocities  preventing   a  reliable  fit  outside  of  this
interval. In order not to rely too  much on the proper motions and mostly on
RAVE  data, the median  proper motion  error of  the sample  being \mbox{1.8
  mas~yr$^{-1}$}, we  first fit only the  constants $A$, $C$ and  $K$ to the
line-of-sight  velocity {\it in  the Galactic  plane} $v_{rad}$,  thus using
only $\vlos$ and  $\mu_b$ .  We then use all  the available information, and
thus fit also  the constants $A$, $B$ and $C$ to  the tangential velocity in
the plane (based on $\mu_l$). Once the Oort constants are fitted to the mean
velocity  field  of  the  observed  population,  we  correct  them  for  the
asymmetric drift  as in OD03: since  the RAVE catalogue is  dominated by the
old thin  disk, we use an asymmetric  drift of 20 km  s$^{-1}$.  The results
are listed in Table~1, and compared to those of KT94 and OD03. We note that,
although we used the updated values of \citet{schoenrich2010} for the reflex
motion  of the  Sun, our  value for  $C$, especially  when fitting  only the
line-of-sight  velocity in the  Galactic plane,  is in  remarkable agreement
with the value found by  OD03 (after the mode-mixing correction) from proper
motions of old red giants\footnote{OD03 found different values for different
  populations  (especially,  lower   $|C|$  for  lower  velocity  dispersion
  populations), but their best tracers of the ``true" Oort constants are the
  red  giants,  also  affected by  an  asymmetric  drift  of roughly  20  km
  s$^{-1}$} in the  ACT/Tycho-2 catalogues.  However, our value  of $C$ (and
$K$) is in disagreement  with the one reported in KT94 from  a review of the
older  literature. Let  us finally  note that  the formal  errors  quoted in
Table~1  were   computed  using  a   Monte-Carlo  sampling  but   are  large
underestimations of  the true errors, due to  possible systematics including
those listed in  OD03, as well as the absence of  a full longitude coverage.
This  underestimation of  the errors  is confirmed  by the  fact  that, when
taking  or not  taking into  account  the longitudinal  proper motions,  the
values of the Oort constants $A$ and $C$ differ by more than 6 formal sigmas
and 4  formal sigmas,  respectively.  The fit  to $v_{rad}$  alone ($3^{rd}$
line  of  Table~1)  is  more  reliable  as it  mostly  relies  on  the  RAVE
line-of-sight velocities  and is  of better quality  than the  one including
longitudinal proper  motions (the  reduced $\chi^2$ values  are 1.4  and 2.6
respectively).

This fitting procedure  of the Oort constants thus  confirmed our finding of
Sect.\ref{s:RV} that $|\partial V_R /\partial R|  \gta 3 \; {\rm km} \, {\rm
  s}^{-1} \,  {\rm kpc}^{-1}$ in the RAVE  catalogue. Let us note  that, as we
rely  on  a   priori  known  distances,  our  result   is  not  affected  by
mode-mixing. However,  in order to check whether  the qualitative conclusion
of the existence  of non-zero velocity gradient depends  on our estimates of
the distances, we assumed that all distances were wrong by a factor $f$, and
re-applied the same procedure. Results for $f=0.8$ and $f=1.2$ are listed at
the end of Table~1, confirming that $K+C \neq 0$ in both cases.

\begin{table*}
\caption{Oort  constants $A$,  $B$,  $C$  and $K$  measured  using the  RAVE
  sample.  The  kinematic centre  $\psi_\nu= 0.5\, {\rm  arctan}(-C/A)$, the
  amplitude  $\sqrt{A^2+C^2}$,  and  the  velocity  gradient  $\partial  V_R
  /\partial R=K+C$ are  also given. All numbers (apart  from $\psi_\nu$) are
  in km~s$^{-1}$~kpc$^{-1}$.  The  first two lines give the  old values from
  KT94 and  OD03.  For  the RAVE  sample, errors bars  are computed  using a
  Monte-Carlo  sampling, but  they do  not  take into  account the  possible
  systematics listed  in OD03. The last  two lines show changes  in the Oort
  constants if one considers a 20\% bias in our distances. The two cases are
  obtained by multiplying  the distances of the whole  catalogue by $0.8$ or
  $1.2$,  then recomputing  the velocities  and re-applying  the  same $R,z$
  selection as before.}
\label{t:oort}
\begin{tabular}{l r c c c c c c c}
\multicolumn{2}{c}{Sample} & $A$ & $B$ & $C$ & $K$ & $\psi_\nu$(deg) & $\sqrt{A^2+C^2}$ & $K+C$\\
\hline
KT94 & $\vlos$, $\mu_b$, and $\mu_l$ & $14.4\pm 1$ & $-12.0 \pm 3$ & $0.6 \pm 1$ & $-0.35\pm0.5$ & -1.2 & 14.4 & 0.25\\
\hline
OD03 & $\mu_l$ & $15.9\pm 2$ & $-16.9 \pm 2$ & $-9.8 \pm 2$ & - & 15.5 & 18.7 & - \\
\hline
 RAVE & $\vlos$ and $\mu_b$ & $13.6\pm0.5$ & - & $-9.6\pm0.5$ & $5.7\pm0.3$ & 17.6 & 16.6 & -3.9 \\
\hline
 RAVE & $\vlos$, $\mu_b$ and $\mu_l$ & $9.2\pm0.5$ & $-17.4\pm0.5$ & $-12.7\pm0.5$ & $4.6\pm0.5$ & 27.0 & 15.7 & -8.1 \\
\hline
$d \times 0.8$ & $\vlos$ and $\mu_b$  & 15.0 & - & -9.2 & 6.3 & 15.8 & 17.6 & -2.9 \\
$d \times 0.8$ & $\vlos$, $\mu_b$ and $\mu_l$ & 11.1 & -11.8 & -10.5 & 5.6 & 21.7 & 15.3 & -4.9 \\
$d \times 1.2$ & $\vlos$ and $\mu_b$  & 12.5 & - & -10.1 & 5.3 & 19.5 & 16.1 & -4.8 \\
$d \times 1.2$ & $\vlos$, $\mu_b$ and $\mu_l$ & 7.9 & -21.3 & -14.4 & 4.1 & 30.6 & 16.4 & -10.3 \\
\hline
\end{tabular}
\end{table*}

%
%

\subsection{Two-dimensional velocity field}
\label{s:2D}

To  further investigate  the velocity  gradient detected  with RAVE,  we now
compute the full  2D velocity field for our samples (making  use of both the
line-of-sight velocities and the proper motion vectors).

The velocity fields in the Galactocentric reference frame are computed using
\begin{equation}
\mathbfss{V}_{GSR}=\mathbfss{V}_{*/\odot}+\mathbfss{V}_{\odot/LSR}
+\mathbfss{V}_{LSR/GSR},
\label{e:basis}
\end{equation}
where $\mathbfss{V}_{\odot/LSR}$  is the Sun's velocity with  respect to the
local standard of  rest (LSR) and $\mathbfss{V}_{LSR/GSR}$ is  the motion of
the LSR with respect to the  GC.  Obviously, a possible radial motion of the
LSR  itself  can be  a  priori  difficult  to disentangle  from  large-scale
streaming  motions.  In  order  to  circumvent this,  we  use hereafter  the
standard assumption  that the LSR  is on a  circular orbit\footnote{Although
  this  does not  formally  exist in  a  non-axisymmetric potential.}.   The
velocity of  stars $\mathbfss{V}_{GSR}$  is hereafter decomposed  into three
components    in   the    axisymmetric   galactocentric    reference   frame
$(V_R,V_{\phi},V_z)$, $V_z$ pointing towards the south Galactic pole.

This  projection depends  on two  parameters :  $R_0$, the  distance  to the
centre of the Galaxy and $v_{c0}$, the local circular speed.  Unfortunately,
as pointed out by \citet{mcmillan2010}, neither of these parameters are well
constrained.  The ratio $v_{c0}/R_0$ is however better constrained, lying in
the  range  $29.9-31.6$~km  s$^{-1}$  kpc$^{-1}$  with some  impact  on  the
measured  $\mathbfss{V}_{\odot/LSR}$   as  shown  in   table  1  and   2  of
\citet{mcmillan2010}.

Therefore, we  compute the 2D velocity  maps for two sets  of parameters. In
both   cases   we   fix   $\mathbfss{V}_{\odot/LSR}$  to   the   values   of
\citet{schoenrich2010}.  The  first set of parameters uses  the standard IAU
values $v_{c0}=220$~km s$^{-1}$ and $R_0=8$~kpc  while for the second set we
use  the  model  2  from table~2  of  \citet{mcmillan2010},  $v_{c0}=247$~km
s$^{-1}$ and  $R_0=7.8$~kpc, which  has a solar  motion vector close  to the
latest determination by \citet{schoenrich2010}.

\begin{figure*}
\centering
\begin{tabular}{p{1.cm} c c}
& $\mathbf{R_0=8}$~{\bf kpc,} $\mathbf{v_{c0}=220}$~{\bf km s}$\mathbf{^{-1}}$& 
$\mathbf{R_0=7.8}$~{\bf kpc,} $\mathbf{v_{c0}=247}$~{\bf km s}$\mathbf{^{-1}}$\\
\begin{sideways}\hspace{2cm}{\bf Full sample}\end{sideways}&
\includegraphics[width=7cm]{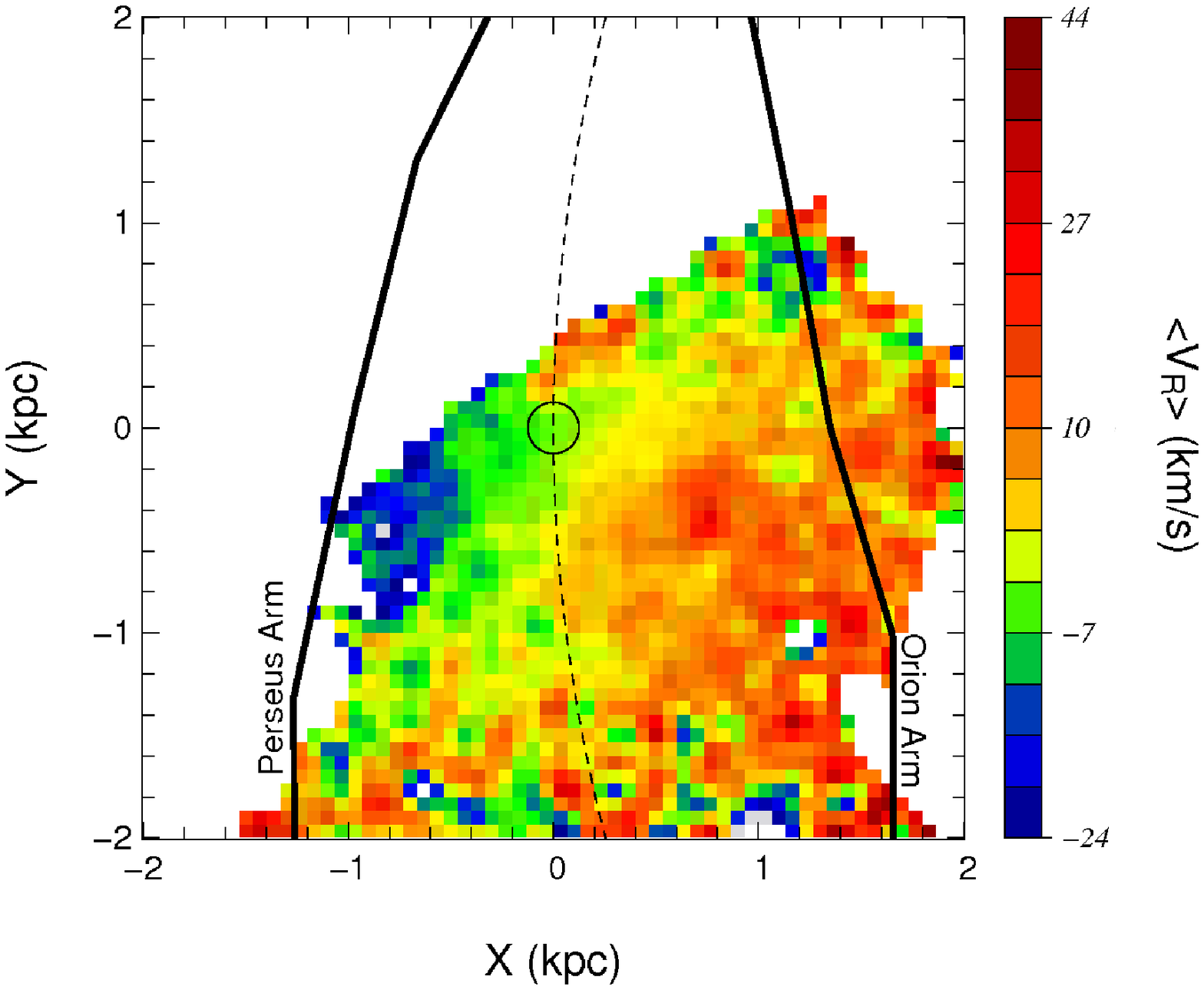}&
\includegraphics[width=7cm]{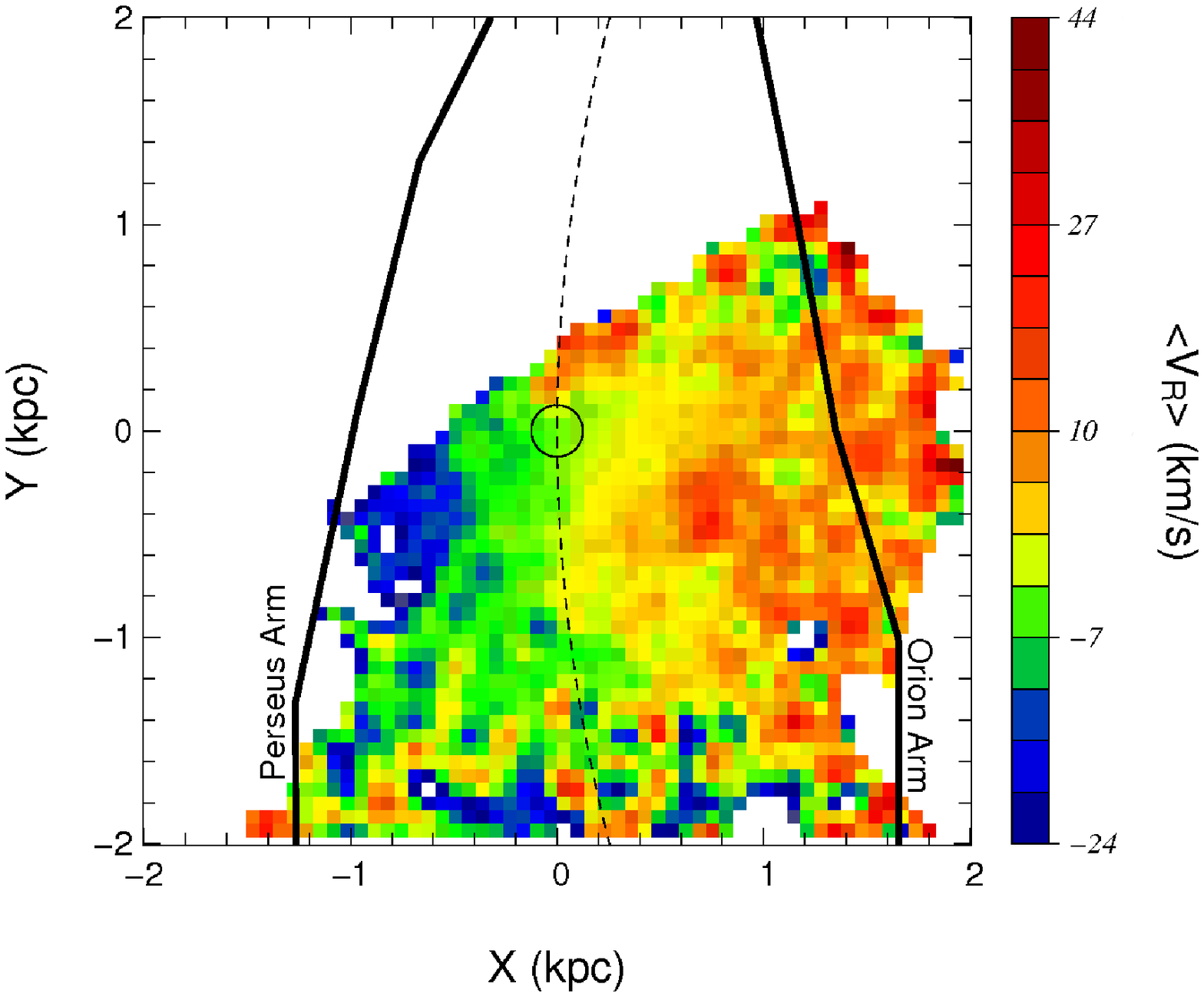}\\
\begin{sideways}\hspace{1.5cm}{\bf Red clump candidates}\end{sideways}&
\includegraphics[width=7cm]{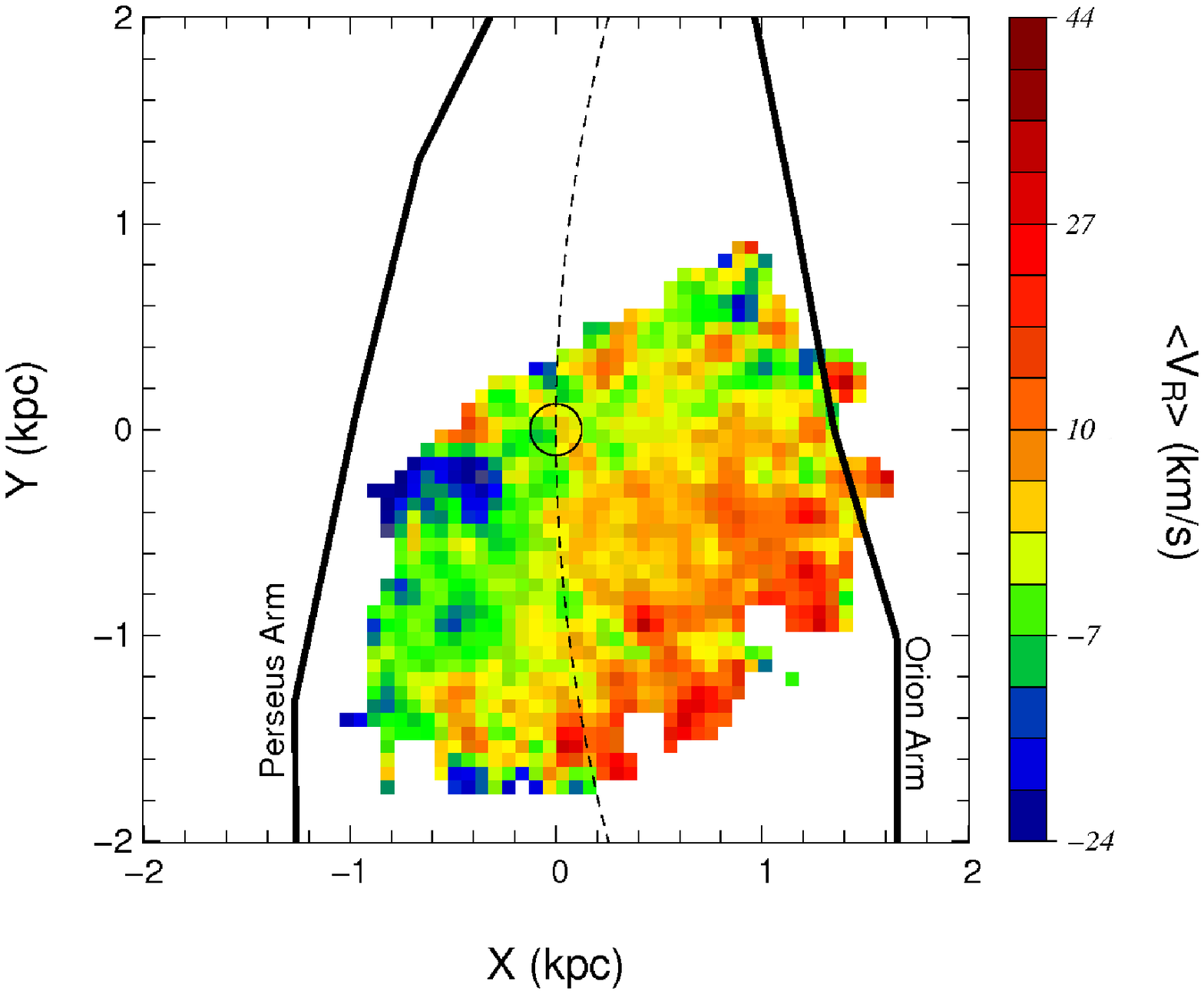}&
\includegraphics[width=7cm]{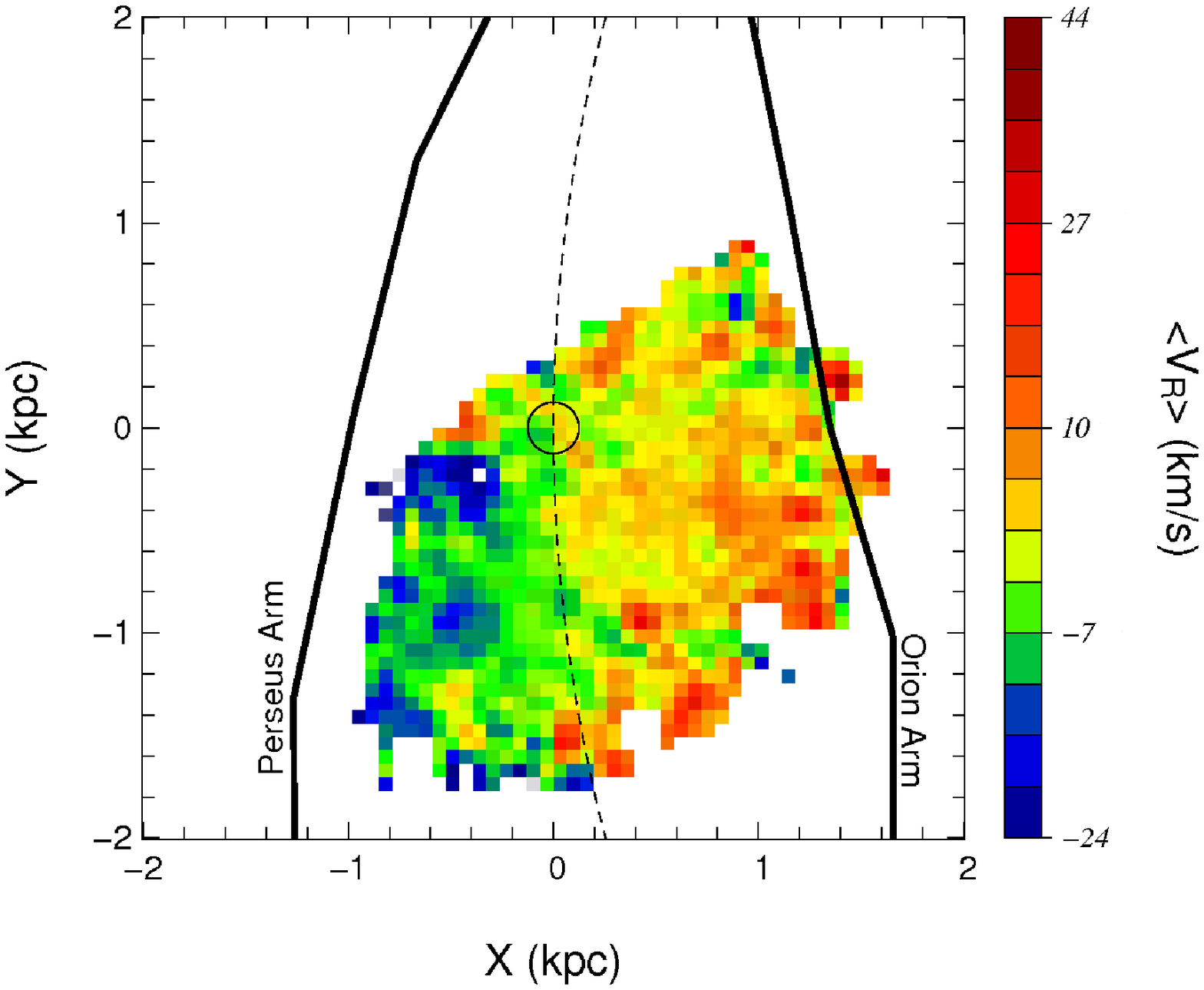}\\
\end{tabular}
\caption{Mean   velocity  fields   $\langle  V_R   \rangle$   derived  using
  Eq.~\ref{e:basis}  and assuming  $R_0=8$~kpc and  $v_{c0}=220$~km s$^{-1}$
  (left  panels)  and  $R_0=7.8$~kpc  and  $v_{c0}=247$~km  s$^{-1}$  (right
  panels).    Top   panels:  full   sample.    Bottom   panels:  red   clump
  selection. The  locations of the  nearest spiral arms are  indicated using
  the CO map  from \citet{englmaier}.  The open circle  delimitates a sphere
  125~pc in radius  around the Sun.  The maps are smoothed  over 3 pixels to
  highlight the mean velocity trends. The maps are $60\times60$ bins in size
  between -2 and 2~kpc along each axis.  X increases in the direction of the
  Galactic centre, Y is positive towards the Galactic rotation.}
\label{f:full}
\end{figure*}

\begin{figure*}
\centering
\begin{tabular}{p{1.cm} c c}
& $\mathbf{R_0=8}$~{\bf kpc,} $\mathbf{v_{c0}=220}$~{\bf km s}$\mathbf{^{-1}}$& 
$\mathbf{R_0=7.8}$~{\bf kpc,} $\mathbf{v_{c0}=247}$~{\bf km s}$\mathbf{^{-1}}$\\
\begin{sideways} \hspace{2cm} {\bf Full sample} \end{sideways} &
\includegraphics[width=7cm]{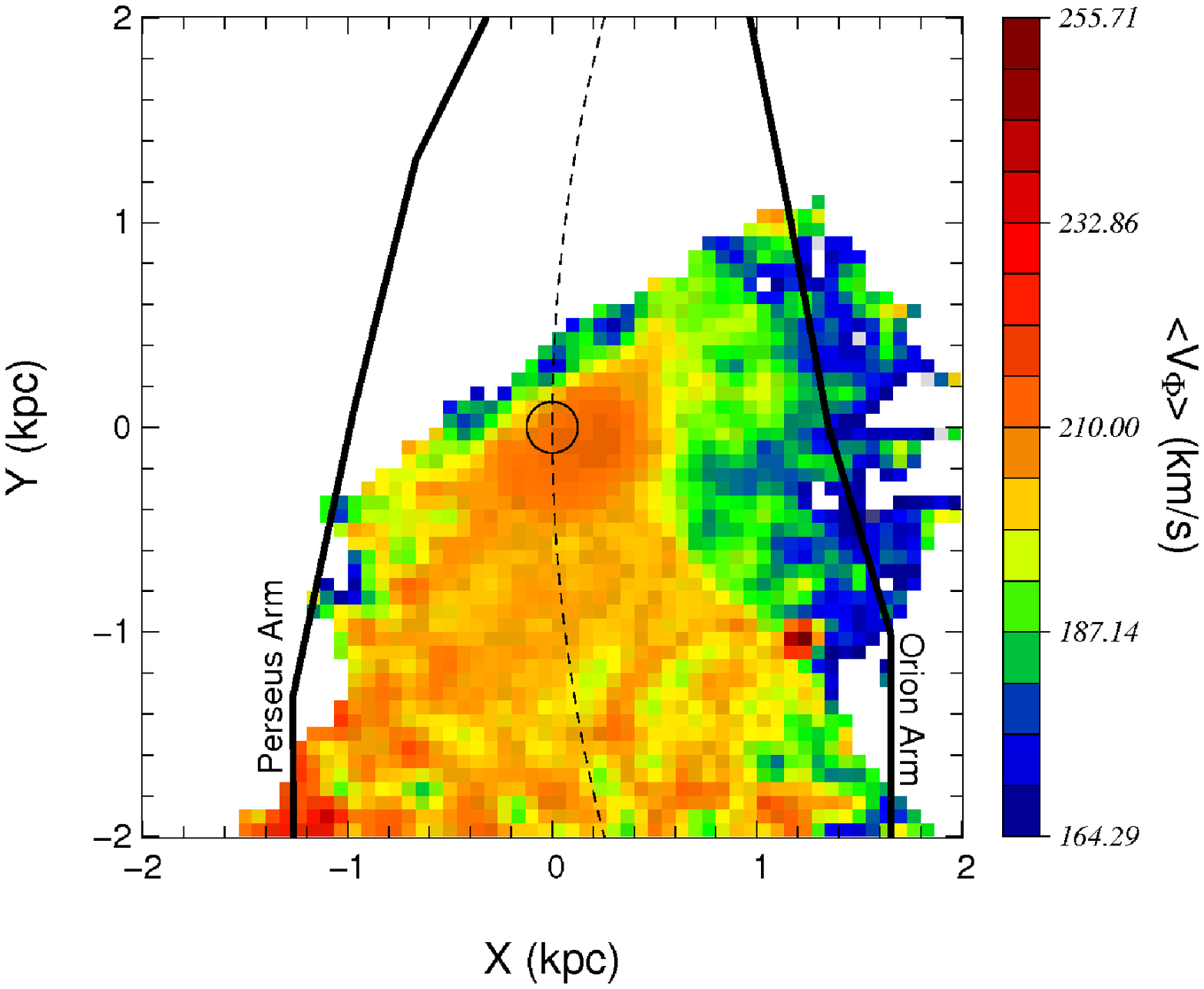}&
\includegraphics[width=7cm]{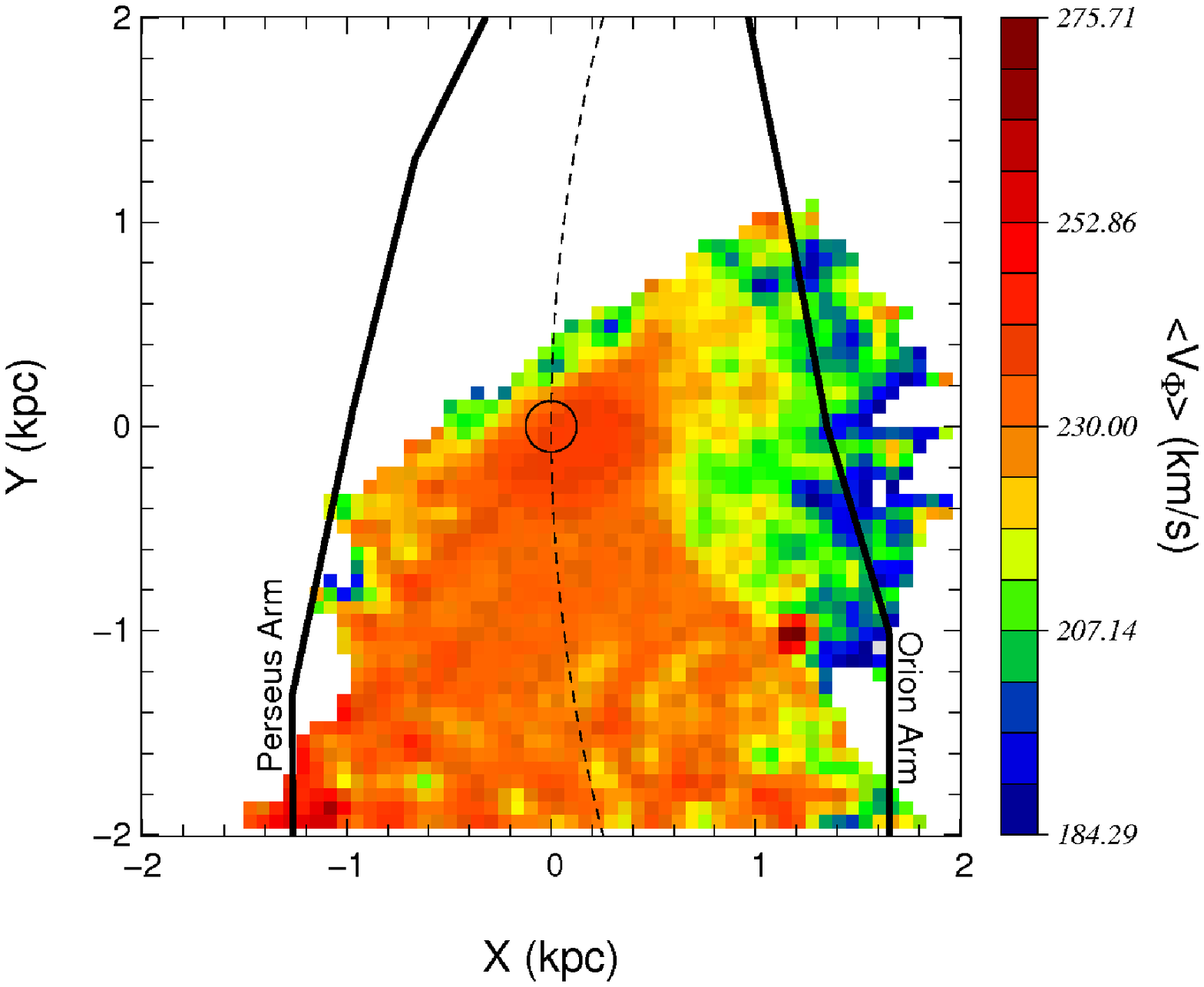}\\
\begin{sideways} \hspace{1.5cm}{\bf Red clump candidates} \end{sideways}&
\includegraphics[width=7cm]{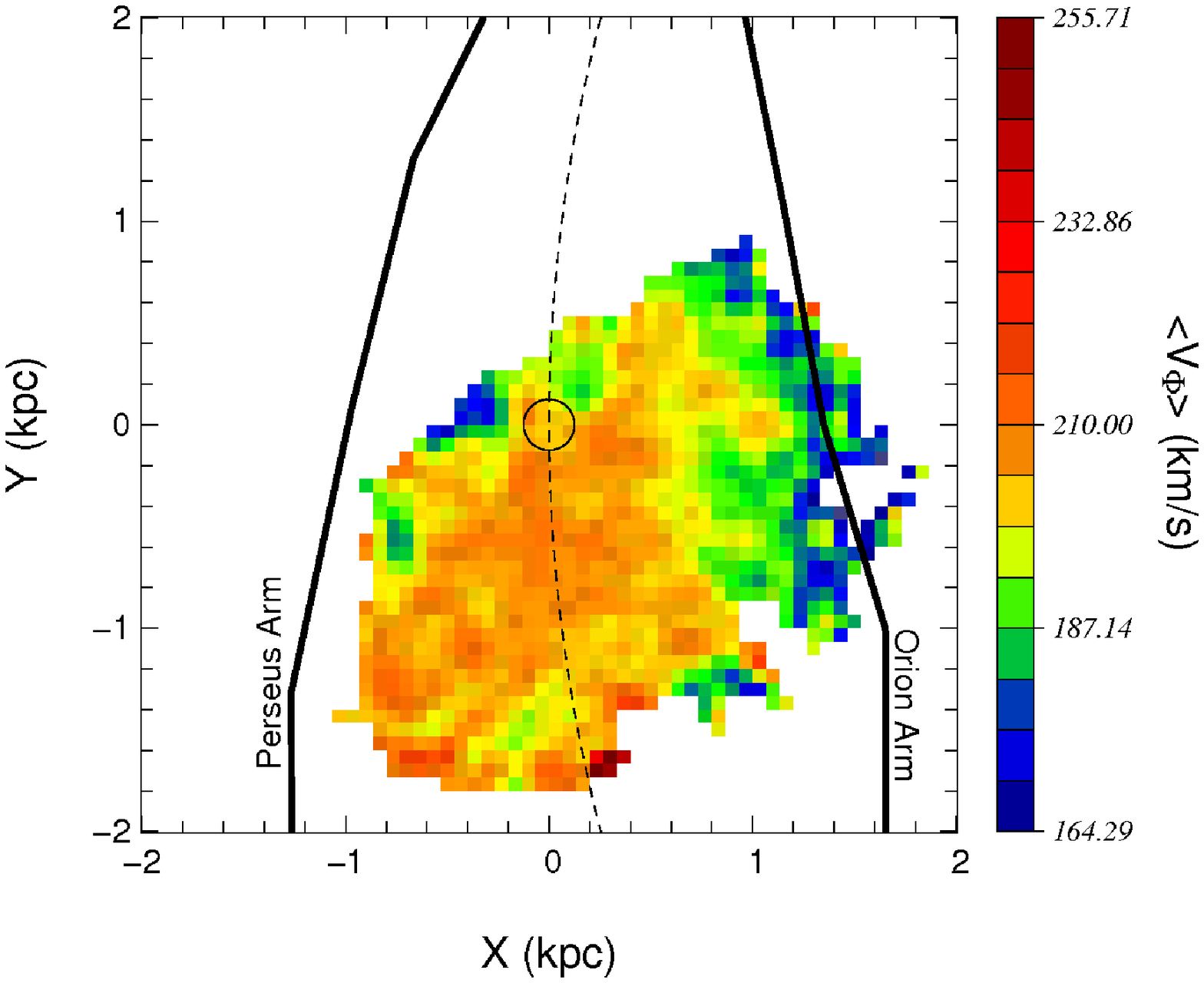}&
\includegraphics[width=7cm]{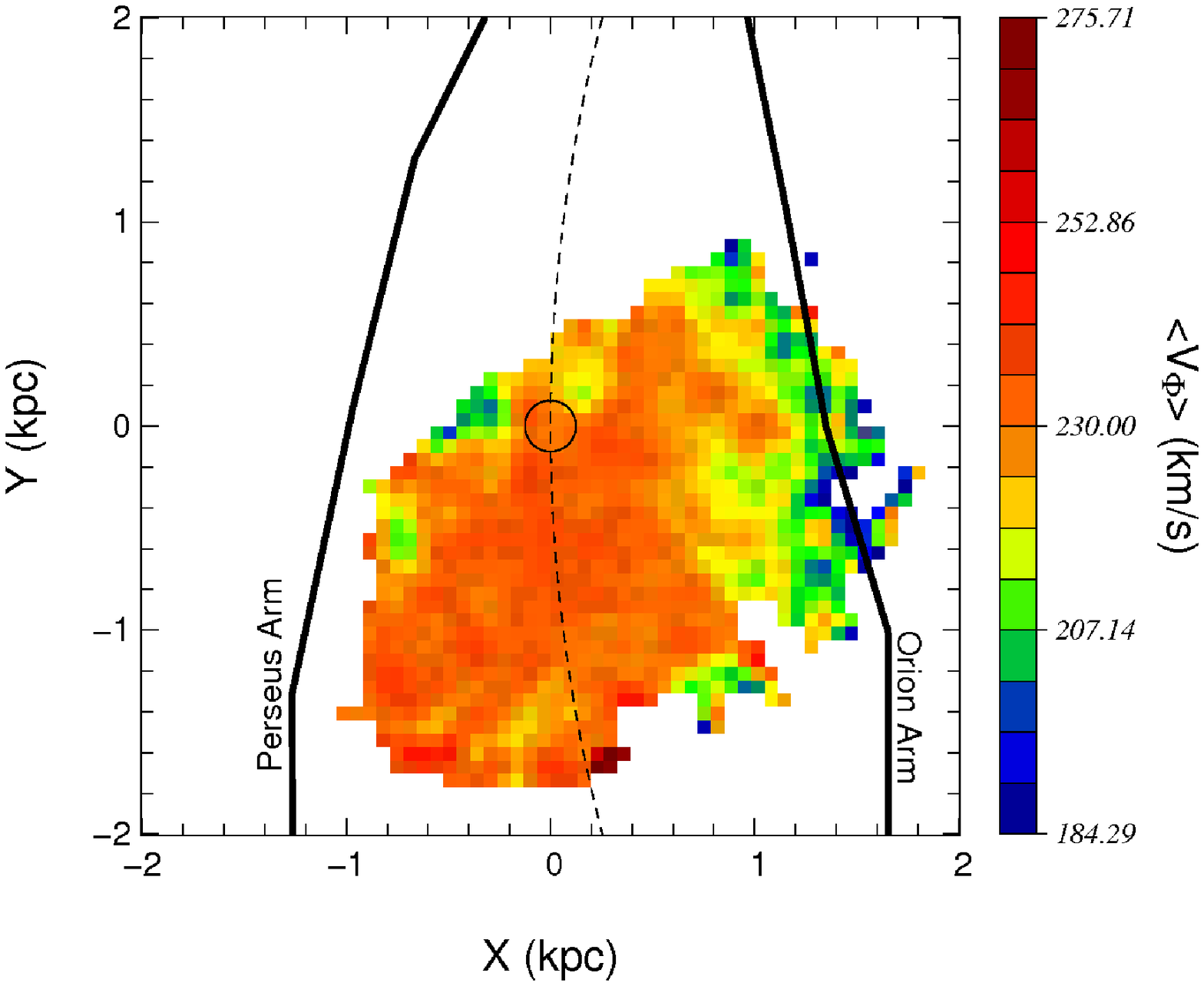}\\
\end{tabular}
\caption{Same as Fig.\ref{f:full} for $\langle  V_\phi   \rangle$.}
\label{f:fullvphi}
\end{figure*}

The results  for the 2D $\langle  V_R \rangle$ and  $\langle V_\phi \rangle$
velocity maps in  the Galactic plane for a  $4\times4$~kpc box ($60\times60$
bins)  centred   on  the  Sun   are  presented  in   Figs.~\ref{f:full}  and
\ref{f:fullvphi}  for both  sets of  parameters and  for both  samples (full
sample and red clump selection).  X and Y increase towards the GC and in the
direction of Galactic  rotation respectively.  This places the  GC either at
(8,0)  or (7.8,0)  depending on  the  Galactic parameters  used.  The  solar
circle ($R=R_0$)  is drawn as a dashed  line and the open  circle delimits a
zone 125~pc in radius similar to the Hipparcos sphere.  For orientation, the
location of the local (Orion) and Perseus arms are indicated based on the CO
map of \citet{englmaier}. No strong kinematic signal is associated with them
in $\langle V_R \rangle$, but  a strong decrease of $\langle V_\phi \rangle$
is associated  with the Orion  arm (while little  data are available  on the
Perseus arm).   This could help  place limits on  the size of  the potential
perturbations associated  with these arms, but  is beyond the  scope of this
paper.

Again, the global structure of  the velocity field in figs.~\ref{f:full} and
\ref{f:fullvphi}, comparing the  top and bottom panels, is  similar for both
samples.   Small  differences  exist  on  local  scales,  especially  around
(X,Y)$=$(0.8,$-0.4$),  but aside from  this region  the velocity  trends are
compatible.  Isocontours  in $\langle {V}_{R} \rangle$  are orientated about
$50\degr$  from the  direction to  the  GC and  do not  follow a  particular
symmetry axis  of our procedure  or sample.  Within the  local neighbourhood
indicated by the  open circle, no net motion is detected  and our values are
compatible  with  previous  measurements.   The  velocity  gradient  becomes
noticeable only outside the Hipparcos sphere.

Comparing left and right  panels of figs.~\ref{f:full} and \ref{f:fullvphi},
we see  that using updated  Galactic parameters does influence  the measured
mean motions.   In particular, in  Fig.~\ref{f:full} it diminishes  but does
not  eliminate the  mean motions.   Trying  to minimise  the gradient  while
keeping the ratio $v_{c0}/R_0$ constant, leads to unrealistic values for the
Galactic parameters with $R_0<4$~kpc and $v_{c0}<126$~km s$^{-1}$.

%
%
\section{Discussion and conclusions} 

In this  paper, making  use of RAVE  line-of-sight velocities  and distances
alone, we report the discovery of a radial velocity gradient in the Galactic
disc  amounting  to  at  least  3~km s$^{-1}$  kpc$^{-1}$  observed  in  the
direction of the GC for $d<2$~kpc.  Making use of the full sample, we fitted
the Oort  constants in  the classical way,  and obtained values  globally in
agreement with those  of OD03 for old red giants.   But as their measurement
was based on proper motions alone, they could not constrain the value of the
Oort constant  $K$, which  we estimate here  to be  roughly between 4  and 6
km~s$^{-1}$~kpc$^{-1}$ (see Table~1).

Then, assuming  a zero  radial velocity  of the Local  Standard of  Rest, we
reconstructed the 2D velocity map, and found that the Galactic parameters do
play  a role in  the measured  amplitude of  the streaming  motion. Adopting
standard  IAU values,  which  are suspected  to underestimate  $v_{c0}/R_0$,
yields a larger amplitude.  On the  other hand, using more recent values for
$v_{c0}$ and $R_0$ reduces the net  radial motion but does not eliminate it.
The  latter  result is  independent  of the  source  of  proper motions  and
distances used.  We note that the confirmed existence of such a net outwards
motion for stars,  amounting to 10\% of the circular  speed, would raise the
question of the validity of determining the Galactic parameters based on the
assumption of  circular motion \citep[see for  example][]{reid2009}, and the
error bars on the measured ratio $v_{c0}/R_0$ are likely underestimated.  We
however also  note that, in  order to establish  it more firmly,  our result
should   imperatively  be  confirmed   by  further   line-of-sight  velocity
measurements at lower latitudes with future spectroscopic surveys \citep[see
  e.g.][]{minchev08}.   It is  indeed very  intriguing that  a  strong 21-cm
absorption feature  along the line-of-sight to  the GC is  compatible with a
zero mean velocity relative to the LSR (Radhakrishnan \& Sarma 1980).

The  cause  of  such a  large-scale  streaming  motion  would clearly  be  a
non-axisymmetric component of  the Galactic potential. It could  thus be due
to: ($i)$ the  Galactic bar; ($ii)$ spiral arms; ($iii)$  the warp; ($iv)$ a
triaxial dark  matter halo; or ($v)$ a  combination of some or  all of these
components.  Indeed, \citet{minchev07a} have  simulated the effect of spiral
density waves  on the  Oort constants and  have showed that  for spiral-only
perturbations,  the value  of $|C|$  is  larger for  lower stellar  velocity
dispersions, contrary  to the  findings of OD03.  The observed trend  in $C$
could,  however,  be   explained  with  the  effect  of   the  Galactic  bar
\citep{minchev07b}, but  not its exact  magnitude. Actually, given  that the
Milky Way  disk contains both spiral arms  and a central bar,  the effect of
both these  perturbers needs to  be considered simultaneously.   In external
galaxies, such  as M81, such non-circular  motions in gas  flows are usually
associated  with spiral  arms and/or  the central  bar. \citet{sellwood2010}
have also placed  bounds on the ellipticity of the  outer dark matter haloes
of  external  galaxies from  the  observed  non-circular motions  \citep[see
  also][]{binney78,franx92,kuijken1994}.  It will be of the highest interest
to determine  which model will  best reproduce the  trend we report  in this
paper.  However, it is unlikely that we will be able to discriminate between
the various  models with  such a  small fraction of  the Milky  Way sampled.
More, deeper observations from  future astrometric and spectroscopic surveys
will probably be mandatory in this respect.

Observations of  non-circular motions of  gas in the inner  Galaxy determine
the characteristics of the central region of the Milky Way and constrain the
parameters of, for  example, the Galactic bar and the  amount of dark matter
allowed      inside      the      solar     circle      \citep[see,      for
  instance,][]{bissantz2003,famaey2005}.   It  is  likely that  mapping  and
understanding the non-circular  motions in the outer parts  of the Milky Way
will reveal similarly important information about the Galactic potential and
the mass distribution on larger scales.

\section*{Acknowledgements}
Funding for RAVE  has been provided by the  Anglo-Australian Observatory, by
the Astrophysical Institute Potsdam,  by the Australian Research Council, by
the German  Research foundation, by the National  Institute for Astrophysics
at  Padova, by  The Johns  Hopkins University,  by the  Netherlands Research
School  for Astronomy,  by  the Natural  Sciences  and Engineering  Research
Council of Canada,  by the Slovenian Research Agency,  by the Swiss National
Science  Foundation,   by  the  National  Science  Foundation   of  the  USA
(AST-0908326), by  the Netherlands Organisation for  Scientific Research, by
the Particle Physics  and Astronomy Research Council of  the UK, by Opticon,
by Strasbourg Observatory, and by  the Universities of Basel, Cambridge, and
Groningen. The RAVE web site is at www.rave-survey.org.



%
\bibliographystyle{aa}

\begin{thebibliography}{}


\bibitem[{{Adler \& Wefstpfahl} (1996)}]{adler1996}
Adler D., Wefstpfahl D., 1996, AJ, 111, 735 

\bibitem[{{Bensby et al.} (2007)}]{bensby07}
Bensby T. et al., 2007, ApJ, 655, L89

\bibitem[{{Binney} (1978)}]{binney78}
Binney J., 1978, MNRAS, 183, 779

\bibitem[{{Binney \& Merrifield} (1998)}]{BM98}
Binney J., Merrifield M., 1998, Galactic Astronomy, Princeton University Press.

\bibitem[{{Binney} (2006)}]{binney2006}
Binney J., 2006, Science, 311, 44

\bibitem[{{Bissantz et al.} (2003)}]{bissantz2003}
Bissantz N., Englmaier P., Gerhard O., 2003, MNRAS, 340, 949 

\bibitem[{{Bosma} (1978)}]{bosma1978}
Bosma, A. 1978, PhD. thesis, University of Groningen 

\bibitem[{{Breddels et al.} (2010)}]{breddels2010}
Breddels M.A. et al., 2010, A\&A, 511, 90

\bibitem[{{Chandrasekhar} (1942)}]{chandra}
Chandrasekhar S., 1942, Principles of Stellar Dynamics, University of Chicago Press.

\bibitem[{{Dame \& Thaddeus} (2008)}]{dame2009}
Dame T.M., Thaddeus P., 2008, ApJ, 683L, 143

\bibitem[{{Dehnen} (1998)}]{dehnen1998}
Dehnen W., 1998, AJ, 115, 2384 

\bibitem[{{Englmaier et al.} (2010)}]{englmaier}
Englmaier P., Pohl M., Bissantz N., 2010, J. Ital. Astron. Soc, in press.

\bibitem[{{Famaey \& Binney} (2005)}]{famaey2005}
Famaey B. \& Binney J., 2005,  MNRAS, 363, 603 

\bibitem[{{Famaey et al.} (2005)}]{fetal05}
Famaey B. et al., 2005, A\&A, 430, 165

\bibitem[{{Franx \& de Zeeuw} (1992)}]{franx92}
Franx M. \& de Zeeuw T., 1992, ApJ, 392L, 47

\bibitem[{{KT94}()}]{kuijken1994}
Kuijken K. \& Tremaine S., 1994, ApJ, 421, 178 (KT94)

\bibitem[{{McMillan \& Binney} (2010)}]{mcmillan2010}
McMillan P.~J., Binney  J., 2010, MNRAS, 402, 934

\bibitem[{{Minchev  \&  Quillen} (2007)}]{minchev07a}
Minchev I. \&  Quillen  A.~C., 2007, MNRAS, 377, 1163

\bibitem[{{Minchev  et al.} (2007)}]{minchev07b}
Minchev I., Nordhaus J.,  Quillen, A.~C., 2007, ApJ, 664, L31

\bibitem[{{Minchev \& Quillen} (2008)}]{minchev08}
Minchev, I. \& Quillen, A.~C., 2008, MNRAS, 386, 1579 

\bibitem[{{Minchev et al.} (2010)}]{minchev2010}
Minchev I., Boily C., Siebert A., Bienaym\'e O., 2010, MNRAS, 407, 2122

\bibitem[{{OD03}()}]{olling03}
Olling R. \& Dehnen W., 2003, ApJ, 599, 275 (OD03)

\bibitem[{{Oort} (1927)}]{oort27}
Oort J.~H., 1927, BAN, 3, 275

\bibitem[{{Pizzella et al.} (2008)}]{pizzella2008}
Pizzella A. et al., 2008, MNRAS, 387, 1099

\bibitem[{{Pont et al.} (1994)}]{pont94}
Pont F., Mayor M., Burki G., 1994, A\&A, 285, 415

\bibitem[{{Quillen \& Minchev} (2005)}]{quillen2005}
Quillen A.~C., Minchev I., 2005, AJ, 130, 576

\bibitem[{{Radhakrishnan \& Sarma} (1980)}]{radha80}
Radhakrishnan V., Sarma N.V.G., 1980, A\&A, 85, 249

\bibitem[{{Reid et al.} (2009)}]{reid2009}
Reid M.~J. et al., 2009, ApJ, 700, 137

\bibitem[{{Rix \& Zaritsky} (1995)}]{rix95}
Rix H.-W., Zaritsky D., 1995, ApJ, 447, 82

\bibitem[{{Sch\"onrich et al.} (2010)}]{schoenrich2010}
Sch\"onrich R., Binney J., Dehnen W., 2010, MNRAS, 403, 1829

\bibitem[{{Sellwood \& Zanmar Sanchez} (2010)}]{sellwood2010}
Sellwood J., Zanmar Sanchez R., 2010, MNRAS in press

\bibitem[{{Siebert et al.} (2008)}]{siebert08}
Siebert A. et al., 2008, MNRAS, 391,793

\bibitem[{{Steinmetz et al.} (2006)}]{dr1} 
Steinmetz M. et al., 2006, AJ, 132, 1645 

\bibitem[{{Xu et al.} (2006)}]{xu2006} 
Xu Y., Reid M. J., Zheng X. W., Menten K. M., 2006, Science, 311, 54

\bibitem[{{Zwitter et al.} (2008)}]{dr2} 
Zwitter T. et al., 2008, AJ, 136, 421

\bibitem[{{Zwitter et al.} (2010)}]{zwitter2010} 
Zwitter T. et al., 2010, A\&A, submitted
\end{thebibliography}


\end{document}